\documentstyle[psfig]{mn}

\def\Rg{{R_{\rm g}}}
\def\Rin{{R_{\rm in}}}

\newcommand\fe{Fe K$\alpha$\ }

\def\dm{{\dot m}}
\def\zycki{$\dot{\rm Z}$ycki}
\def\lubinski{Lubi\'nski}

\newcommand\fx{F_{\rm x}}
\newcommand\fdisk{F_{\rm disk}}
\newcommand\rozanska{R\'o$\dot{\rm z}$a\'nska }
\def\simlt{\lower.5ex\hbox{$\; \buildrel < \over \sim \;$}}
\def\simgt{\lower.5ex\hbox{$\; \buildrel > \over \sim \;$}}
\def\sqr#1#2{{\vcenter{\hrule height.#2pt
      \hbox{\vrule width.#2pt height#1pt \kern#1pt \vrule width.#2pt}
         \hrule height.#2pt}}}

\title[X--ray irradiated disks]
{Testing models of X--ray reflection from irradiated disks}
\author[C. Done \& S. Nayakshin]
{C. Done$^1$ \& S. Nayakshin$^{2,3}$ \\
$^1$ Department of Physics, University of Durham, South Road,
Durham, DH1 3LE, England; chris.done@durham.ac.uk \\
$^2$ USRA and LHEA, NASA/Goddard
Space Flight Center, Greenbelt, MD20771, USA;
serg@milkyway.gsfc.nasa.gov\\
$^3$ Max Planck Institut fur Astrophysik,
Karl-Schwarzschild-Str. 1,
Postfach 1317, D-85741 Garching, Germany}

\begin{document}

\maketitle

\begin{abstract}

We model the reflected spectrum expected from localized magnetic
flares above an ionised accretion disk. We concentrate on the case of
very luminous magnetic flares above a standard accretion disk
extending down to the last stable orbit, and use a simple
parameterisation to allow for an X--ray driven wind.  Full disk
spectra including relativistic smearing are calculated.  When fit with
the constant density reflection models, these spectra give both a low
reflected fraction and a small line width as seen in the hard spectra
from Galactic Black Hole Binaries and Active Galactic Nuclei. We fit
our calculated spectra to real data from the low/hard state of Nova
Muscae and Cyg X--1 and show that these models give comparable
$\chi^2$ to those obtained from the constant density reflection models
which implied a truncated disk.  This explicitally demonstrates that
the data are consistent with either magnetic flares above an ionized
disk extending down to the last stable orbit around a black hole, or
with non-ionized, truncated disks. 

\end{abstract}


\section{INTRODUCTION}

X--ray reflection is {\em in principle} a sensitive diagnostic of the
X--ray source geometry. The amount of reflection depends on the solid
angle subtended by the optically thick material, (and also on the
ionisation and elemental abundances in the reflector) while the amount
of relativistic smearing of the reflected spectral features
(especially the iron K$\alpha$ line and edge) shows how far this
material extends down into the gravitational potential (see the review
by Fabian et al. 2000 and references therein).  This is important as
there is much current debate about the structure of the X--ray source.
One of the possibilities discussed in the literature is magnetic
reconnection above the disk (e.g. Galeev, Rosner \& Vaiana 1979,
Haardt, Maraschi \& Ghisellini 1994) while another is the advection
dominated accretion flows (e.g. Narayan \& Yi 1995).  These result in
very different source geometries -- in the advective flows the inner
disk is truncated, being replaced by an X--ray hot plasma, while the
magnetic reconnection models have a disk which extends down to the
last stable orbit.

The reflected spectrum is generally calculated {\it assuming} a given
density structure of the disk (generally constant) as a function of
height, and with a constant ionisation state as a function of
radius. Fits with these models to spectra from 
both Galactic Black Hole Candidates
(GBHC) and some Active Galactic Nuclei (AGN) show overwhelmingly that
the solid angle subtended by the disk is generally less than unity,
and that the relativistic smearing is less than expected for a disk
which extends down to the last stable orbit (\zycki, Done \& Smith 1997;
1998; 1999; Zdziarski, \lubinski\ \& Smith 1999; Chiang et al.,
2000; Done, Madejski \& \zycki\ 2000; Miller et al., 2001). 
They also show that the solid angle and amount of smearing
are {\it correlated} with spectral index in the
sense that harder spectra show a smaller amount of reflection
($\Omega/2\pi <<1$) and narrower line emission (\zycki\ et al., 1999;
Zdziarski et al., 1999; Gilfanov, Churazov \& Revnivstev
1999; 2000; \lubinski\ \& Zdziarski 2001). 
Ionisation of the reflected spectrum is
only strongly present for steep spectrum sources. 

This rules out the simplest version of the magnetic flare model unless
the flares generate a mildly relativistic outflow away from the disk
(Beloborodov 1999). However, such outflows need not be required if the
modelled reflected spectra are systematically underestimating the
amount of reflection present in the data. For example, if the inner
disk is so highly ionised that even iron is completely stripped, then
reflection produces no spectral features (e.g. Fabian et al., 2000 and
references therein).  However, the data require a fairly sharp
transition between the extreme ionisation and mainly neutral material,
so as not to produce an intermediate region of high ionization, where
iron is mainly H-- and He--like, which produces huge and unmistakable
spectral features which are {\it not} present in the data (Done \&
\zycki\ 1999; Done et al., 2000; Young et al., 2001).

It has recently been realised that X--ray illumination {\it changes}
the density structure of the disk. The X--ray heated material at the
top can expand outwards, lowering its density and increasing its
ionisation. Calculating the self consistent density (and hence
ionization) structure is especially important as there is a thermal
ionization instability which affects X--ray illuminated material in
pressure balance, which can lead to a hot, extremely ionised skin
forming on top of the rest of the cooler, denser, mainly neutral disk
material (Field 1965; Krolik, McKee \& Tarter 1981; Kallman \& White
1989; Ko \& Kallman 1994; \rozanska\ \& Czerny 1996; Nayakshin,
Kazanas \& Kallman 2000, hereafter NKK; \rozanska\ \& Czerny 2000;
Nayakshin \& Kallman 2001, hereafter NK; Ballantyne, Ross \& Fabian
2001).

The existance of the thermal instability can cause a substantial
underestimate of the amount of reflection present as measured by the
fixed density reflection model spectra (NKK; NK).  This is explicially
shown by Done \& Nayakshin (2001, hereafter DN) and Ballantyne et al.,
(2001) for reflection from a single radius in the disk.  
Here we extend the calculations of DN to build
full disk reflection models for the case of magnetic flares.  We give
a simple parameterization of the one major uncertainty in these
models, which is the strength of any sideways expansion of the X--ray
heated material from underneath the flare (see \S 2). We first fit the
calculated total spectra (incident spectrum plus reflection from a
disk with self consistent density structure 
including the thermal instability) with the power law plus fixed
density reflection models in \S 3. We then fit the new models to
actual data in \S 4 and show for the first time that the magnetic
flare models (in which the disk subtends a solid angle of
$2\pi$) can fit the observed apparently weak reflection/narrow line, 
flat spectrum GBHC rather well. However, with these models,
the data {\it require} moderate outflow rates from the X--ray heated
region.  Future work should explicitly calculate this sideways
expansion of material, so as to remove the one remaining free
parameter from the simplest magnetic flare model.

\section{REFLECTED SPECTRA}

The code of NKK gives the reflected spectrum from an X--ray
illuminated slab whose vertical structure is given initially by that
of an accretion disk (Shakura \& Sunyaev 1973)
at radius, $r$, and dimensionless
mass accretion rate, $\dot{m}=\eta \dot{M} c^2/{\rm L_{Edd}} = {\rm L}
/{\rm L_{Edd}}$ (${\rm L_{Edd}}$ is the Eddington accretion rate, $L$
is the disk bolometric luminosity, $\dot{M}$ is the mass accretion
rate and $\eta= 0.06$ is the radiative efficiency of the standard disk
in Newtonian limit), where a fraction $f_{corona}$ of the energy
liberated by this infalling material is dissipated in a corona rather
than in the optically thick disk (Svensson \& Zdziarski 1994). This
also defines the soft flux $\fdisk$ at the top of the disk due to
viscous dissipation underneath. There is also additional heating from
the illuminating X--ray flux $\fx$ with spectral shape $\fx(E)$, which
then changes the vertical structure since the heated upper layer can
expand. The steady state density structure is calculated under the
hydrostatic equilibrium assumption.

The reflected spectrum from a full disk is an integral of reflected
spectra from different radii. In practice we compute spectrum for only
6 radii due to CPU time limitations: $r=$~3.5, 4.9, 7, 14, 35 and 105
Schwarzchild radii for the same $\dot{m}$ and X--ray spectral shape,
with a given radial dependance of the X--ray illuminating flux which
depends on the X--ray source geometry.  If the X-ray source(s) are
magnetic flares due to the MHD dynamo responsible for the viscosity
(Balbus \& Hawley 1991) then the X--ray flux might be expected to
scale with the disk flux i.e. $F_x\propto F_{disk}\propto r^{-3}$ so
that $F_x/F_{disk}$ is constant. The spectra from different radii can
then be convolved with the relativistic transfer functions (assuming a
Schwarzchild metric, but not including light bending: Fabian et
al. 1989), which yields the requisite full disk relativistically
broadened spectra.

Given the four initial parameters ($\dot{m}$, $f_{corona}$,
$F_x/F_{disk}$, and X--ray spectral shape), together with the
assumption that $F_x\propto F_{disk}$, the reflected spectra from
magnetic flares should be completely determined for a given
inclination angle. However, the assumption of hydrostatic equilibrium
is {\it not} entirely justified because a local outflow is formed
(NKK, NK).  Hydrostatic balance predicts that the vertical scale
height for a skin of temperature $T$ is $H_s \sim (4 kT
R^3/GMm_p)^{1/2}$. Compare this with the standard disk vertical
pressure scale height, $H$, using equations of Svensson \& Zdziarski
(1994):
\begin{equation}
\left(\frac{H_s}{H}\right)^2 = 2.5\times 10^{-3} T_7
\left(\frac{\dm(1-f_{corona})}{0.01}\right)^{-2}
r^3 J(r)^{-2}\;,
\label{hs1}
\end{equation}
where $T_7$ is the skin's temperature in units of $10^7$ Kelvin, 
$\dm$ is the dimensionless accretion rate, $r\equiv Rc^2/2GM$, and
$J(r) = 1 - \sqrt{3/r}$. At radius $r=6$, for example, this ratio
becomes
\begin{equation}
\left(\frac{H_s}{H}\right)^2 = 6.5\, T_7 
\left(\frac{\dm(1-f_{corona})}{0.01}\right)^{-2}\;.
\label{hs2}
\end{equation}
For Cyg~X-1, $\dm(1-f)$ is likely to be less than 0.01, and
$T_7\simgt$ few for hard X-ray spectra. Now, magnetic flares are
likely to have height not much larger than $\sim H$ (Nayakshin \&
Kazanas 2001b) above the disk. Hence, the hydrostatic balance condition
predicts that the skin can extend to height $H_s \gg H$, i.e.,
completely surround the X-ray source. This is inconsistent
with the explicit assumption of the NKK code that X-rays are incident
on the top of the skin at a given angle (equivalent to the source
located at infinity from the disk). 

In addition, when the X-ray source is engulfed by the highly
photo-ionized skin, there is a large radiation pressure force directed
away from the X-ray source. The gas expands upwards (away from the
disk) and sideways (along the disk) due to the combination of thermal
pressure and radiation pressure. The gas is moved on distances $H_s\gg
H$ before the radiation force and illumination decreases so that the
gas can both cool and fall back onto the disk.  Note that the material
{\em cannot} escape to infinity as it is gravitationally bound -- the
escape temperature from the inner accretion disk is $T\sim 10^{11}$ K
(Begelman, McKee \& Shields 1983).  Note also that ``local'' winds do
not occur for other than the magnetic flare models (see, e.g., NKK;
NK, \S 4.1), where the hydrostatic balance assumption should be a good
approximation.

NKK approximated the effects of this local outflow by introducing a
``gravity parameter'' $A={\cal F}_g/{\cal F}_{rad}$, which is the
ratio of the vertical component of the gravitational force at the top
of the disk, ${\cal F}_g = GMH\rho/R^3$, to the radiation pressure
force, ${\cal F}_{rad}= \fx\sigma_t n_e /c$.  For a given $\fx$,
$\dot{m}$ and $f_{corona}$, the gravity parameter is uniquely
determined. However, when the wind is present, hydrostatic balance
does not apply as we discussed above. Therefore, NKK and NK allowed
$A$ to be a free parameter (even when $\fx$ is fixed) to account for
the poorly constrained strength of the wind.  The Thomson thickness of
the skin, $\tau_s$, should always be lower if the wind is present
compared with the values predicted by hydrostatic balance. By
increasing the gravity parameter artificially, one also obtains
smaller values of $\tau_s$, hence qualitatively one can account for
the wind by making $A$ larger than its true value.

Here, instead of artificially increasing the gravity parameter, $A$,
we try to develop a simple approximation to a local wind that should
place us in approximately the correct part of parameter space. In particular,
we write
\begin{equation}
\frac{\partial p}{\partial z} = - \frac{p}{\lambda}\;.
\label{eq2}
\end{equation}
where $p$ is the gas pressure, and $\lambda$ is the characteristic
scale on which the gas pressure is decreasing. If hydrostatic balance
assumption were appropriate, this scale would be the usual thermal
pressure scale height (e.g., see $\lambda$ as defined in Nayakshin
2000). Here the geometry leads us to consider the pressure changing on
scales of order the height of the flare, which is of order $H$
(Nayakshin \& Kazanas 2001b). The coefficient of
proportionality can only be found in a full scale 2 or 3-D calculation
of this problem. Simple estimates show it can depend on the ratio of
the X-ray and disk fluxes, $\fx/\fdisk$, the Thomson thickness of the
skin, disk accretion rate, etc. Therefore, we introduce a free
dimensionless ``wind parameter'', $\Lambda$, to describe this
uncertainty: $\lambda \equiv \Lambda H$. 

Simple estimates show that $\Lambda \sim 1$ is appropriate to a
thermally driven wind when the gas density decreases significantly at
the distance $\sim H$ of the flare location. The case of small
$\Lambda$ corresponds to supersonic outflows, $v\gg c_s$ ($c_s$ is the
sound speed), when $\lambda \sim H c_s/v$. We will vary $\Lambda$
below and study its influence on the reflected spectra.

Summarizing, we use equation (\ref{eq2}), together with
$\lambda=\Lambda H$, to replace the hydrostatic balance condition in
the NKK code (eq. 19 in NKK). The lower boundary conditions --
continuity in the gas density and $T=T_{\rm eff}$ remain the same.
The code then directly computes the density structure of the X--ray
irradiated disk including the effects of this outflow, so this is
clearly an improvement on the previous description which used
hydrostatic balance with the {\em ad hoc} gravity parameter, $A$.  In
terms of gas temperature profiles, the results are not appreciably
different than those using $A$ in that the 
nearly discontinuous transition between the hot
and cold phases is still present. However, the advantage of the new
approach is that we hope
(\ref{eq2}) automatically puts us in the roughly the correct parameter
space for real magnetic flares.

\section{SIMULATED SPECTRA}

We simulate spectra over a range of expected parameters for the hard
X--ray spectra seen from GBHC and AGN. Plainly the spectral index of
the incident power law influences the results, so we test two values
of the index: $\Gamma=1.6$ and $\Gamma=1.8$. The power law is assumed
to extend out to $200$ keV, and then have an abrupt cutoff. This is a
better model of a Compton scattered spectrum below $\sim 80$ keV than
the very gradual rollover produced by an exponential
cutoff. Observations seem to indicate $f_{corona} \sim 0.8$, so we fix
this in all simulations.  We assume an underlying disk accretion rate
of $\dot{m}=0.03$ (this contrasts with the $\dot{m}=10^{-3}$ assumed
in paper 1) onto a super--massive black hole of $10^8 {\rm M}_\odot$.
The large black hole mass means that we avoid disk photons potentially
contributing to the observed bandpass, and more importantly
numerically we avoid the very high densities in GBHC disks for which
the atomic data and codes are not reliable. XSTAR 2.0 (Kallman \&
Bautista 2001) has the highest density range of current
photo--ionisation codes, but even this breaks down at $\sim 10^{18}$
cm$^{-3}$, whereas the GBHC disks can have densities of $\sim 10^{20}$
cm$^{-3}$ in the ionised skin, and $10^{22}$ cm$^{-3}$ in deeper
layers. It is simply not possible to calculate the ion populations in 
X--ray illuminated GBHC disks with current codes. 
We use the AGN reflected spectra as the best current
approximation to those from GBHC, but we caution that the GBHC ion
populations can be different from those given here.

\begin{figure*}
\centerline{\psfig{file=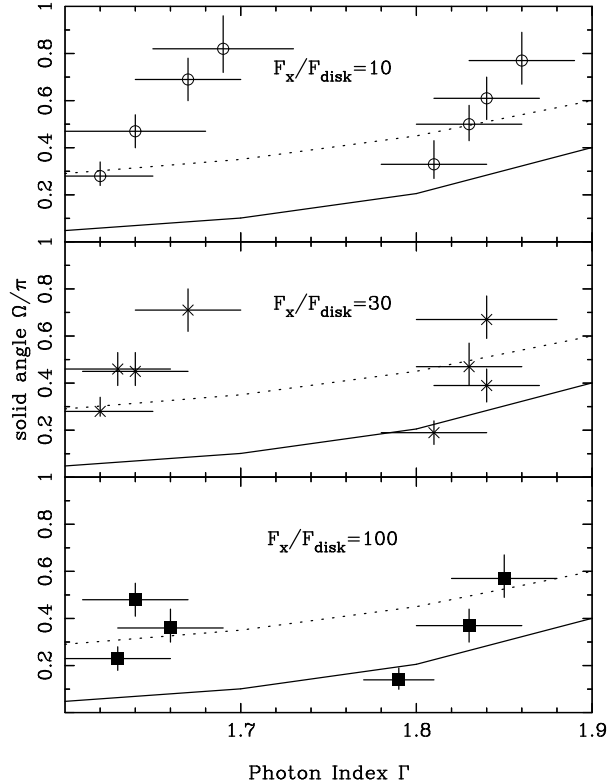,width=0.45\textwidth}}
\caption{Derived values of spectral index and reflected fraction for 
simulated RXTE datafiles made from the illuminated disk models,
fit with the {\tt pexriv} constant density reflection model.  The
models are calculated for $F_x/F_{disk}=10$ (large circles; upper
panel), $30$ (crosses; middle panel),
and $100$ (squares; lower panel), 
each with wind parameter $\Lambda=0.03, 0.1, 0.3$
and $1$ (except for $F_x/F_{disk}=100$ where $\Lambda=0.03,0.1$ and
$0.3$). Stronger winds (small $\Lambda$)
have smaller optical depth in the extremely ionized skin, and so have
larger reflected fractions. The solid line in each panel is the observed
index--reflection correlation of Zdziarski et al. (1999), while the
dotted line is that observed for GBHC (Gilfanov et al.,
2000).}

\end{figure*}

To obtain flat spectra, i.e. $\Gamma < 2$, 
requires that the power dissipated in heating
the electrons in the flare 
is much larger than the seed photon luminosity available
for Compton scattering (e.g., Pietrini \& Krolik 1995; Stern et al. 1995,
Poutanen \& Svensson 1996).
Further, in order to produce reflected spectra with little
He-like Fe contribution, one needs the skin to be essentially
completely ionized. Within a factor of few, both of these conditions
require the X-ray flux incident on the disk surface near flares
footpoints to be much larger than the disk flux, i.e., $\fx\gg
\fdisk$.  We thus decided to test values of $\fx/\fdisk$ anywhere
between $\sim 10-100$, and we choose a grid of $\fx/\fdisk=10, 30,
100$.  We also choose a grid for the wind parameter of $\Lambda=0.03,
0.1, 0.3$ and $1$.

This results in 24 model spectra of reflection from a disk which
extends down to the last stable orbit. We simulate these spectra
through the RXTE PCA response, assuming a flux of $\sim 10^{-8}$ ergs
cm$^{-2}$ s$^{-1}$ and a 1ks exposure (i.e. comparable to a single PCA
orbit of a bright GBHC). The resulting 3--20 keV spectra have
systematic errors of $0.5$ per cent added in quadrature to the
(negligible) statistical errors since it is sadly unlikely that any
satellite will be calibrated to better accuracy than this.  These
spectra are fit using the fixed density ionization models without
relativistic smearing ({\tt pexriv}: Magdziarz \& Zdziarski 1995),
with a separate narrow (intrinsic width fixed at $0.1$ keV) Gaussian
line of free normalization, and energy (except that this latter is
constrained to be between 5.5 and 7.5 keV, as expected for iron with
some shifts allowed for Doppler and gravitational effects).  All the
fits were statistically acceptable ($\chi^2_\nu=0.6-1.2$ for 38
degrees of freedom) and 
gave a derived ionization consistent with
mostly neutral material except for the case with $\fx/\fdisk=100$,
$\Lambda=1$. This extreme illumination with slow wind
results in a large optical depth in the hot layer ($\ge 1$), so the
observed reflected spectrum comes from the lower temperature material
in the hot layer rather than the cool material underlying the skin.
This gives rise to a clearly ionised reflected signature, so 
these data are not shown in the following plots. 

Figure 1 shows the resulting derived photon spectral index, $\Gamma$,
and solid angle, $\Omega/2\pi$ for each different $\fx/\fdisk$. 
For a given spectral index, the
optical depth of the ionized layer is larger for higher $\fx/\fdisk$,
and for slower winds. The solid line shows the overlaid
$\Gamma-\Omega/2\pi$ correlation derived by Zdziarski et al., (1999)
for a sample of AGN and GBHC, while the dotted line shows that derived
from GBHC by Gilfanov et al., (2000). Clearly, the simulated spectra
can match the observed correlations only if the wind strength and
$\fx/\fdisk$ are varied in the ``right way''. Unfortunately values of
these parameters depend crucially on the (currently unknown) physics
of magnetic flares, e.g. on how the height of the flare might evolve
as a function of mass accretion rate, so it is hard to see whether
these parameters will indeed cooperate to produce the observed
correlations.

We get similar results (see Figure 2) in terms of derived spectral
index and reflected fraction when using the reflection code developed
by \zycki\ et al., (1999). This code has the iron line intensity and
energy tied self--consistently to the continuum reflection, and then
both line and reflected continuum are smeared by relativistic effects.
Again the models fit the simulated data well, and imply a low
ionization state of the reflecting material, but the difference now is
that the spectral broadening of the line and reflected edge can be
constrained. Figure 2b shows the amount of smearing (in units of
gravitational radii $\Rg=GM/c^2$ rather than Schwarzchild radii,
$R_s=2\Rg$). In general a higher optical depth (slower
wind and higher $\fx/\fdisk$) gives less smearing, but {\it none} of
the spectra show the extremely smeared and skewed components which are
associated with the last stable orbit around the black hole.

\begin{figure*}
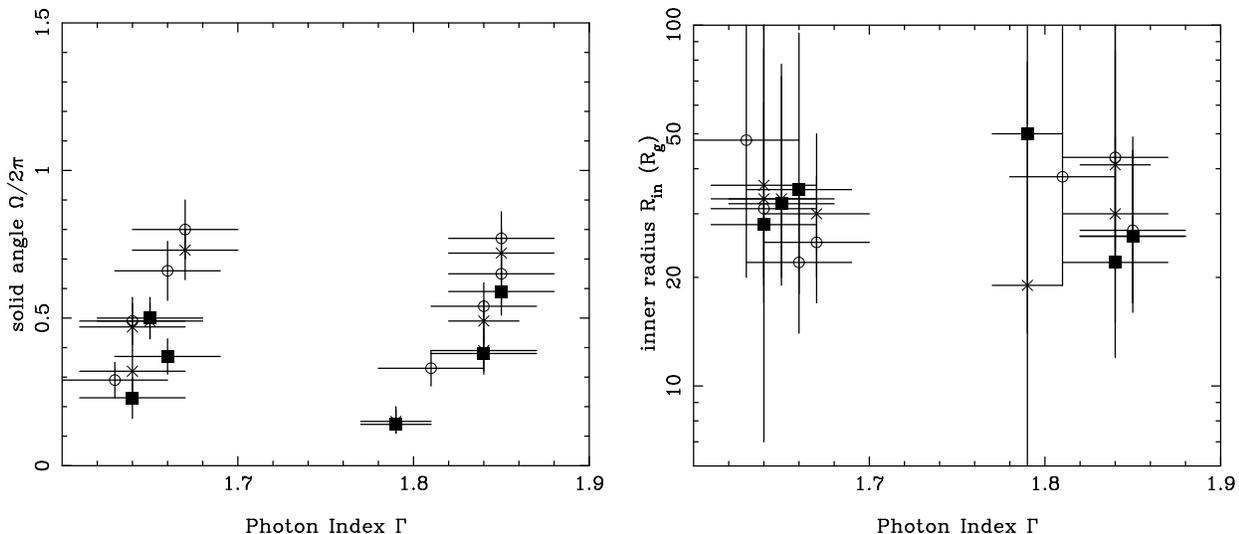

\begin{tabular}{cc}
{\psfig{file=fig2a.ps,width=0.45\textwidth}}
&
{\psfig{file=fig2b.ps,width=0.45\textwidth}}
\end{tabular}
\caption{(a) Derived values of spectral index and 
reflected fraction from the simulated XTE datafiles shown in Figure 1
but fit with constant density reflection models of \zycki\ et al.,
(1999). These include the self--consistently calculated iron emission
lines for a given reflected continuum, and relativistically smear both
line and continua together. The symbols are the same as for Figure
1. It is again clear that the optical depth of the extremely ionized
skin is the major factor which determines the properties of the
observed spectrum, but this is dependent on the (currently unknown)
wind strength.  (b) Derived values of the inner disk radius in units
of $\Rg=GM/c^2$ obtained from the fits in (a), plotted against the
spectral index. While the RXTE resolution is fairly poor,
the line is {\it never} as broad as expected if the reflector were a
constant density disk extending down to the last stable orbit in a
Schwarzchild geometry ($\equiv 6\Rg$).}
\end{figure*}

\begin{figure*}
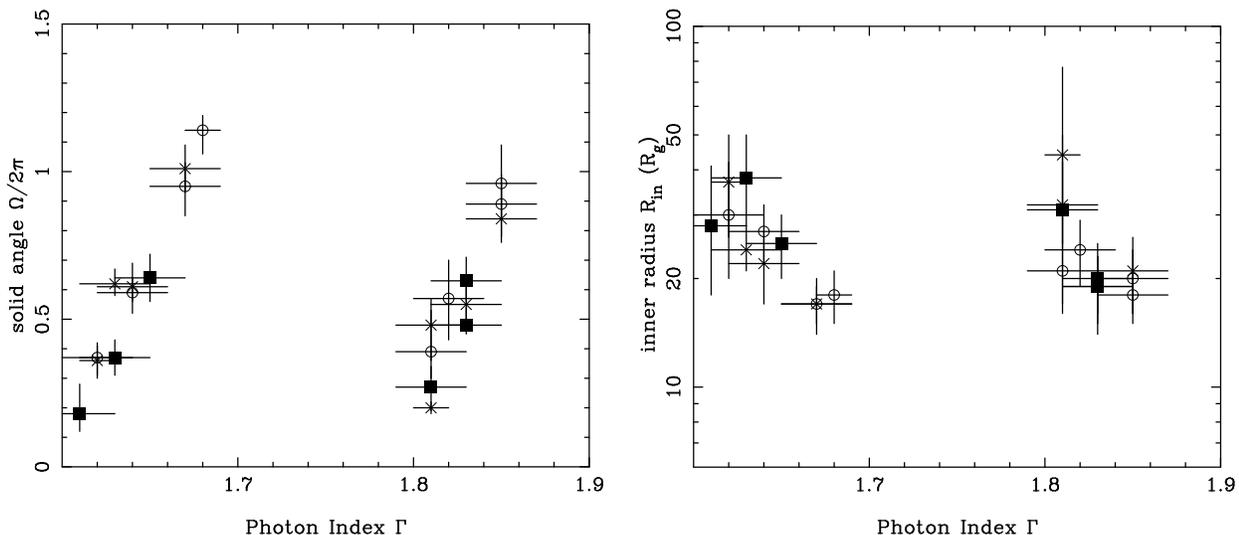

\begin{tabular}{cc}
{\psfig{file=fig3a.ps,width=0.45\hsize}}
& 
{\psfig{file=fig3b.ps,width=0.45\hsize}}
\end{tabular}
\caption{As for Figure 2a and b, but with fits to 
simulated ASCA GIS datafiles. (b) The 
moderate resolution of the GIS detectors gives better constraints 
on the amount of relativistic smearing than those obtained from the 
RXTE simulations (figure 2b). Plainly the ionized skin 
suppresses the line emission from the innermost regions, leading to derived
disk radii which are {\it always} larger than 3 Schwarzchild radii
($\equiv 6 \Rg$).
}
\end{figure*}

Clearly the resolution from proportional counters leaves much to be
desired. To investigate the smearing further we simulated the model
spectra through the ASCA GIS response, assuming a mean 2--10 keV flux
of $\sim 10^{-8} $ ergs cm$^{-2}$ s$^{-1}$, and a 20 ks exposure (as
typical for the Cyg X--1 observations: Ebisawa et al., 1996).  We fit
these spectra with the relativistically smeared reprocessed spectral
model of \zycki\ et al., (1999) over the 4--10 keV band (for
comparison with the GIS Cyg X--1 data of Done \& \zycki\ 1999).  Again
all the fits are adequate ($\chi^2_\nu=0.75-1.25$ for 79 degrees of
freedom) with the reflected spectrum ionisation indicating 
mainly neutral material.
Figure 3a shows the resulting derived spectral index and solid angle,
showing that curvature in the data can be more easily matched in the
narrower bandpass data by a somewhat steeper spectral index and
correspondingly higher reflected fraction. Figure 3b shows the
resulting inner disk radii. Again the reflected spectral features are
{\it never} as broad as expected if the reflector were a constant
density disk extending down to the last stable orbit in a
Schwarzchild geometry.

This latter fact is somewhat surprizing given that the Thomson depth
of the skin is quite low, $\tau_s\sim 0.1$, for the strongest wind
models. Experimenting with fits, we found that the main reason that
the inner disk radius is larger in the fits than in the actual models
is that the broad redshifted wing of the \fe line gets confused with
the incident power-law. Indeed, the inferred value of $\Gamma$ differs
from the actual one by as much as $0.1$ for the strongest wind
cases.

\section{REAL DATA}

\begin{figure*}
\centerline{\psfig{file=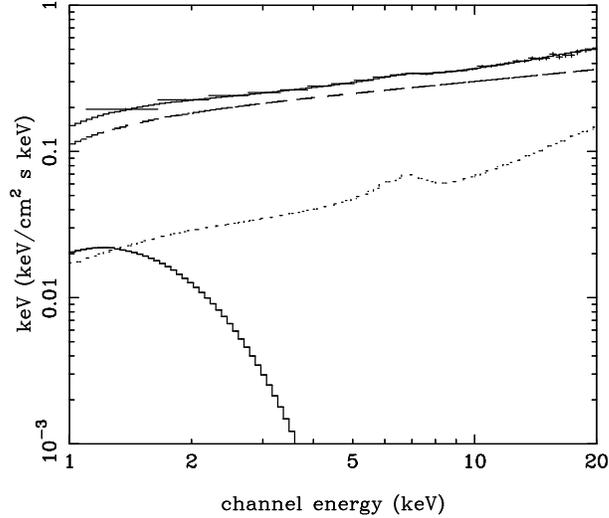,width=0.45\textwidth}}
\caption{Data from the transient black hole Nova Musca taken on July
23 1991 when the source showed a classic low state spectrum. The data
are well described by the complex ionization reflection models
presented here.
The dashed line shows the incident power law (which is absorbed at
low energies by the interstellar column of $1.6\times 10^{21}$
cm$^{-2}$) while the dotted line shows its reflection from a disk with
a completely ionized skin ($\Lambda=0.3$, $F_x/F_{disk}=10$,
$i=63^\circ$). The lower solid line shows the contribution from the
disk blackbody spectrum.}
\end{figure*}

\begin{figure*}
\centerline{\psfig{file=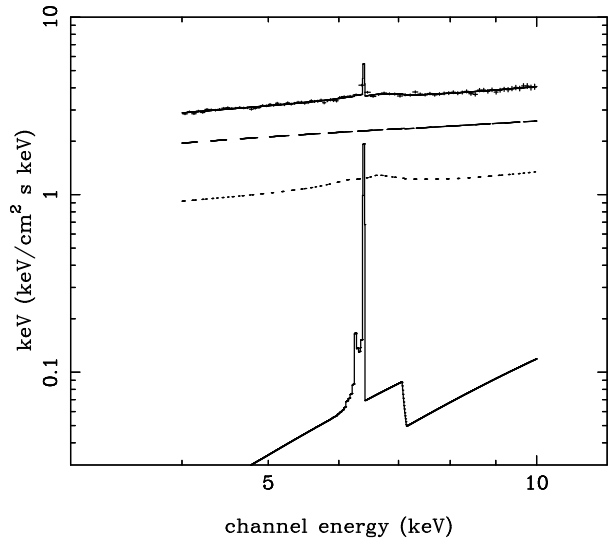,width=0.45\textwidth}}
\caption{ASCA GIS data for Cyg X--1 fit to the 
complex ionization reflection models presented here.
The dashed line shows the incident power law 
while the dotted line shows its reflection from a disk
with a completely ionized skin ($\Lambda=0.3$, $F_x/F_{disk}=100$,
$i=49^\circ$). The lower solid line shows the unsmeared, neutral
reflection from the companion star/outer disk.}
\end{figure*}

Plainly these model spectra bear a close resemblance to the data
in terms of the properties of the reflected spectrum. The next step is
to fit such models to real data. As a first attempt at this we
have taken the tabulated spectra used above and incorporated them into
the XSPEC spectral fitting package (Arnaud 1996) as a local model.
With this approach the only true free parameter is the normalisation of
the reflected spectrum, but we have approximated the change made by
small variations of spectral slope ($\Delta\Gamma
\le 0.1$ by multiplying the models calculated for a given spectral
index $\Gamma_0$ by the factor $E^{\Gamma-\Gamma_0}$. We have models
tabulated for $\Gamma_0=1.6$ and $1.8$, and these give 
{\it reflected} spectrum for $\Gamma=1.7$ which are consistent to 
within 5\% in the 2--20 keV band, so the total spectra for
$\Gamma=1.7$ derived from $\Gamma_0=1.6$ and $1.8$ are consistent to
within 2 \%. 

The first test is for broad band data from GINGA. A classic GBHC low
state spectrum is that of Nova Muscae on July 23rd 1991.
The constant density (single ionization parameter) reflection
models give a best fit of $\chi^2=15.3/24$ for a model consisting of a
disk blackbody, together with a power law and its relativistic
reflection component (inclination fixed at $60^\circ$) with $\Gamma=1.72\pm
0.02$, $\Omega/2\pi=0.24^{+0.11}_{-0.08}$,
$\Rin=50^{+\infty}_{-35}\Rg$, all absorbed by a (fixed) column of 
$N_H=1.6\times 10^{21}$ cm$^{-2}$ (\zycki\ et al., 1998). 
We replace the power law and its relativistic
reflection by the model spectra used above (incident flux plus its
reflection from the complex ionization structure of the disk) for
$\Gamma_0=1.8$ and step over all tabulated values of $F_x/F_{disk}$ and
$\Lambda$. The best fit is shown in Figure 4, giving 
$\chi^2_\nu=18.5/25$ for $F_x/F_{disk}=10,
\Lambda=0.3$. Plainly low resolution, broad band data 
where the reflected spectrum appears to have much lower normalisation
and smearing than expected for a disk around a black hole
can be well fit by these model spectra which have the disk subtending
a solid angle of unity and extending down to the last stable orbit.

A higher resolution spectrum gives a better test of the relativistic
smearing constraints. We use the ASCA GIS 6 data from Cyg X--1
of Ebisawa et al.,
(1996), as analysed by Done \& \zycki\ (1999) in the 4--10 keV range.
We step over inclinations from $56^\circ-31^\circ$ as the inclination
of the Cyg X--1 system is not well known, and include a narrow,
neutral reflected component (solar abundance, inclination of
$60^\circ$, with line normalisation and energy tied to the reflected
continuum) to account for the contribution to the reflected spectrum
from the companion star. The interstellar column is fixed at $6\times
10^{21}$ cm$^{-2}$. This gives a best fit of $\chi^2_\nu=83.8/79$ for
$\Lambda=0.3, F_x/F_{disk}=100, i=49^\circ$, shown in Figure 5.  There
are many other solutions within $\Delta\chi^2=6.25$ (90 per cent
confidence for 3 parameters), although none of these allow the
strongest wind with $\Lambda=0.03$. The quality of the fit is plainly
comparable to that of the constant ionization models of Done \&
\zycki\ (1999), who obtained $\chi^2_\nu=81.4/79$, 
$\Omega/2\pi=0.10^{+0.17}_{-0.09}$ and $\Rin=18^{+30}_{-12}\Rg$.  Thus
even high resolution data which appears to show much less reflection
and (marginally) smearing from the accretion disk than expected can be
fit with the models where the ionization instability gives rise to a
completely ionized skin on the disk overlying more neutral material.

\section{CONCLUSIONS}

We use a modified version of the code of NKK and NK to calculate the
vertical structure of an X--ray illuminated disk at a given radius,
and then sum these spectra over all radii, with appropriate
relativistic smearing, to get the full disk reflected spectra expected
from magnetic flares.  However, the reflected spectra are {\it not}
uniquely determined as the local intense illumination expected from
magnetic flares can drive outflows from the illuminated region,
lowering the optical depth of the ionized skin. We approximate the
effects of this with a free parameter $\Lambda$ (equation \ref{eq2}).
We simulate full disk reflected spectra for a grid of different
$F_x/F_{disk}=10, 30$ and $100$, with spectral index $\Gamma=1.6, 1.8$
and with $\Lambda=0.03,0.1,0.3$ and $1$.

We simulate each total spectrum (illuminating power law plus full disk
reflection) through the low resolution RXTE 
and moderate resolution ASCA GIS response and fit these with
the constant density (single ionization state) {\tt pexriv} models.  The
ionized skin reduces the derived $\Omega/2\pi$ for a given spectral
index, and also reduces the amount of relativistic smearing observed
as the ionized skin depth is at its maximum in the inner disk.

The resemblance of the (low--to moderate wind velocity) simulated spectra to
observations of low--state GBHC and AGN is striking. The models have
the disk sub--tending a solid angle of $2\pi$ and extending down to the
last stable orbit, but the vertical ionization structure can mask the
reflected signature especially from the inner disk.  When such spectra
are fit with the constant density reflection models then the derived
solid angle and iron line width {\it appear} to imply that the disk
subtends a small solid angle and has a large inner disk radius. These
simulations of a full disk confirm the conclusions of DN and
Ballantyne et al., 2001, drawn
from local (single radius) models that the observations of a small
solid angle and line width do not necessarily rule out static magnetic
flares above an untruncated disk.

We show for the first time that these magnetic flare models can fit
real data. Both low resolution (GINGA data from the low/hard state of
Nova Muscae, \zycki\ et al., 1998) and moderate resolution (ASCA GIS
data from the low/hard state of Cyg X--1, Done \& \zycki\ 1999) can be
fit with these full disk models. The $\chi^2$ for these fits is comparable
to that obtained from the constant density reflection models which
implied a truncated disk. Thus the data can be consistent with {\it
all} the currently proposed X--ray production mechanisms -- either
simple (static) magnetic flares, or with outflowing plasma from
a magnetic flare, or with a truncated disk/inner X--ray hot (advective) flow.

To make progress we clearly need more sophisticated models of the
reflected spectra. It is probable that the primary reason that all
these different models can fit the data is because of the current
freedom in the values of model parameters (e.g., $\Lambda$ for the
wind from magnetic flares: the relativistic outflow velocity in the
plasma ejection models: Beloborodov 1999; the transition radius in the
advective flow models: Esin, McClintock \& Narayan 1997).  In
particular, the magnetic flare models should be improved via an
explicit calculation of the X--ray induced wind. This may limit the
range of possible $\Lambda$ for a given $\fx/\fdisk$, ruling some
models out. For the cold outer disk plus inner hot flow models, the
next step is to include the presence of the skin on the cold disk --
the physics of the thermal ionization instability applies to all
configurations of the disk.  Another modelling issue which needs to be
addressed is the current inability of the atomic codes to handle the
high densities required to properly model the GBHC disks.

We also need better data, especially observations which extend over a
wider bandpass. Including the higher energy spectra beyond 20~keV may
start to break the degeneracy between the different model spectra
(DN), while simultaneous lower energy data can constrain the seed
photon flux.  Another possible way to distinguish between the models
is to study their variability behaviour. The ionised skin of the disk
can respond to changes in the illuminating flux on a timescale of the
order of the Keplerian rotation time scale. This time scale is longer
than the light crossing time at the same radius (Nayakshin \& Kazanas
2001a). In other words, a highly variable light curve may produce
reflected spectra that are different from the steady-state models
(studied in this paper), which then could be used to differentiate
between the models. Finally, and perhaps most convincingly, \fe line
reverberation studies should present a variety of constraints on the
models (Reynolds 2000, Revnivtsev et al. 1999). Recently,
Nayakshin \& Kazanas (2001b) showed that time-resolved \fe line from a
single magnetic flare is {\em narrow} but is moving across the $\sim
4-8$ keV band. This prediction is completely different from the
instantaneous line profile expected in a central source or
lamppost--like geometry (e.g., Young \& Reynolds 2000, Ruszkowski
2000), and one can hence hope that that will be the definitive test of
the accretion flow geometry.

\section{ACKNOWLEDGEMENTS}

SN acknowledges support from NRC Research Associateship. The authors
thank Demos Kazanas and Piotr \zycki\ for useful discussions. We are
indebted to the anonymous referee for pointing out a very serious
typographical error in the equation (\ref{eq2}).

{}


\begin{thebibliography}{}


\bibitem[]{} Arnaud K. A., 1996, Astronomical Data Analysis Software and
Systems V, eds. Jacoby G. and Barnes J., p17, ASP Conf. Series volume 101. 

\bibitem[]{} Balbus S. A., Hawley J. F., 1991, ApJ, 376, 214

\bibitem[]{} Ballantyne D., Ross R., Fabian A. C., 2001, MNRAS, in press





\bibitem[]{} Begelman M. C., McKee C., Shields G., 1983, ApJ, 271, 70

\bibitem[]{} Beloborodov A. M. 1999, ApJ, 510, L123


\bibitem[]{} Chiang J., Reynolds C. S., Blaes O. M., Nowak M. A., Murray N.,
Madejski G., Marshall H. L., Magdziarz P., 2000, ApJ, 528, 292

\bibitem[]{} Done C., Madejski G. M., \zycki\ P. T., 2000, ApJ, 536, 213

\bibitem[]{} Done C., Nayakshin S., 2001, ApJ, 546, 419 (DN)

\bibitem[]{} Done C., \zycki\ P. T., 1999, MNRAS, 305, 457

\bibitem[]{} Ebisawa K., Ueda Y., Inoue H., Tanaka Y., White, N. E., 1996, 
ApJ, 467, 419

\bibitem[]{} Esin A. A., McClintock J. E., Narayan R., 1997, ApJ, 489, 865

\bibitem[]{} Fabian A. C., Rees M. J., Stella L., \& White N. E.,
1989, MNRAS, 238, 729

\bibitem[]{} Fabian A. C., Iwasawa K., Reynolds C. S., Young A., 2000, 
PASP, 112, 1145

\bibitem[]{} Field G. B., 1965, ApJ, 142, 431

\bibitem[]{} Galeev A. A., Rosner R., Vaiana G. S., 1979, ApJ, 229, 318



\bibitem[]{} Gilfanov M., Churazov E., Revnivtsev M., 1999, A\&A,
352, 182

\bibitem[]{} Gilfanov M., Churazov E., Revnivtsev M., 2000, 
5th CAS/MPG Workshop on High Energy Astrophysics (astro-ph/0002415)


\bibitem[]{} Haardt F., Maraschi L., Ghisellini G., 1994, ApJ, 432, 95

\bibitem[]{} Kallman T. R., Bautista M., 2001, ApJS, 133, 221

\bibitem[]{} Kallman T. R., White N. E., 1989, ApJ, 341, 955

\bibitem[]{} Ko Y.--K., Kallman T. R., 1994, ApJ, 431, 273

\bibitem[]{} Krolik J. H., McKee C. F., Tarter C. B., 1981, ApJ, 249, 422



\bibitem[]{} \lubinski\ P., Zdziarski A. A., 2001, MNRAS, 323, L37

\bibitem[]{} Magdziarz P., Zdziarski A. A., 1995, MNRAS, 273, 837

 

 

\bibitem[]{} Miller J. M., Fox D. W., Di Matteo T., Wijnands R.,
Belloni T.,  Pooley D.,  Kouveliotou C., Lewin W. H. G., 2001, ApJ, 546,
1055

\bibitem[]{} Narayan R., Yi I., 1995, ApJ, 444, 231

\bibitem[]{} Nayakshin S. 2000, ApJ, 534, 718.

\bibitem[]{} Nayakshin S., \& Kallman T. R., 2001, ApJ, 546, 406 (NK)

\bibitem[]{} Nayakshin S., \& Kazanas D., 2001a, ApJ, 553, L141

\bibitem[]{} Nayakshin S., \& Kazanas D., 2001b, submitted to
ApJ

\bibitem[]{} Nayakshin S., Kazanas D., \& Kallman T. R., 2000, ApJ,
537, 833 (NKK)

\bibitem[]{} Pietrini P., Krolik J. H., 1995, ApJ, 447, 526


\bibitem[]{} Poutanen J., Svensson R., 1996, ApJ, 470, 249

\bibitem[]{} Revnivtsev M., Gilfanov M., \& Churazov E., 1999, A\&A,
347, L23

\bibitem[]{} Reynolds C. S., 2000, ApJ, 533, 811




\bibitem[]{} \rozanska\ A., Czerny B., 1996, AcA, 46, 233

\bibitem[]{} \rozanska\ A., Czerny B., 2000, A\&A, 360, 1170

\bibitem[]{} Ruszkowski M., 2000, MNRAS, 315, 1    

\bibitem[]{} Shakura N. I., Sunyaev R. A., 1973, A\&A, 24, 337

\bibitem[]{} Stern B. E., Poutanen J., Svensson R., Sikora M., Begelman
M. C., 1995, ApJ, 449, 13

\bibitem[]{} Svensson R., Zdziarski A. A., 1994, ApJ, 436, 599

\bibitem[]{} Young A., Reynolds C. S., 2000, ApJ, 529, 101

\bibitem[]{} Young A., Fabian A. C., Ross R., Tanaka Y., 2001, MNRAS,
in press

\bibitem[]{} Zdziarski A. A., Lubi\'nski P., Smith D. A., 1999, MNRAS,
303, 11


\bibitem[]{} \zycki\ P. T., Done C., Smith D. A., 1997, ApJ, 488, L113
 
\bibitem[]{} \zycki\ P. T., Done C., Smith D. A., 1998, ApJ, 496, L25
 
\bibitem[]{} \zycki\ P. T., Done C., Smith D. A., 1999, MNRAS, 305, 231

\end{thebibliography}
\end{document}